\documentclass[%
 reprint,
superscriptaddress,
 amsmath,amssymb,
 aps,
]{revtex4-2}

\usepackage{graphicx}
\usepackage{dcolumn}
\usepackage{bm}
\usepackage{xcolor}
\usepackage{graphicx}
\usepackage{amsmath}
\usepackage{siunitx}

\begin{document}

\preprint{APS/123-QED}

\title{Double resonant tunable second harmonic generation in two-dimensional layered materials through band nesting}


\author{Sudipta Romen Biswas$^\dagger$}
\affiliation{Department of Electrical \& Computer Engineering, University of Minnesota, Minneapolis, MN 55455, USA}
\author{Jin Yu$^\dagger$}
\affiliation{Shanghai Key Laboratory of Mechanics in Energy Engineering, Shanghai Institute of 
	Applied Mathematics and Mechanics, School of Mechanics and Engineering Science, 
	Shanghai University, Shanghai, 200444, China}
\affiliation{Institute for Molecules and Materials, Radboud University, Heijendaalseweg 135, 
	6525AJ Nijmegen, The Netherlands}
\author{Zhenwei Wang}
\affiliation{Key Laboratory of Artificial Micro- and Nano-Structures of Ministry of Education and 
	School of Physics and Technology, Wuhan University, Wuhan 430072, China}
\author{Diego Rabelo da Costa}
\affiliation{Department of Electrical \& Computer Engineering, University of Minnesota, Minneapolis, MN 55455, USA}
\affiliation{Departamento de F\'isica, Universidade Federal do Cear\'a, Campus do Pici, 60455-900, Fortaleza, CE, Brazil}
\affiliation{Key Laboratory for Micro/Nano Optoelectronic Devices of Ministry of Education \& Hunan Provincial Key Laboratory of Low-Dimensional Structural Physics and Devices, School of Physics and Electronics, Hunan University, Changsha 410082, China}
\author{Chujun Zhao}
\affiliation{Key Laboratory for Micro/Nano Optoelectronic Devices of Ministry of Education \& Hunan Provincial Key Laboratory of Low-Dimensional Structural Physics and Devices, School of Physics and Electronics, Hunan University, Changsha 410082, China}
\author{Shengjun Yuan}
\affiliation{Key Laboratory of Artificial Micro- and Nano-Structures of Ministry of Education and 
	School of Physics and Technology, Wuhan University, Wuhan 430072, China}
\affiliation{Institute for Molecules and Materials, Radboud University, Heijendaalseweg 135, 
	6525AJ Nijmegen, The Netherlands}
\author{Tony Low \thanks{tlow@umn.edu}}
\affiliation{Department of Electrical \& Computer Engineering, University of Minnesota, Minneapolis, MN 55455, USA}
\affiliation{School of Physics and Astronomy, University of Minnesota, Minneapolis, MN 55455, USA}

%

\begin{abstract}
We proposed a mechanism to generate giant anisotropic second harmonic nonlinear response via double resonance effect, achieved through band nesting via electronic bandstructure engineering. The ideal band setup would be a triplet of nested bands separated by the fundamental resonance energy, $\hbar\omega$. We demonstrate theoretically that the proposed phenomenon can be realized in bilayer SnS by band tuning with perpendicular electrical bias, which maximizes the second harmonic susceptibility by several orders of magnitude. Moreover, the tunability of the polarization anisotropy can be useful for realizing novel polarization-sensitive devices.
\end{abstract}

\maketitle


\section{Introduction}
When the electric field intensity is high (on the order of 100 kV/m or more), the materials' response to the field would acquire a notable nonlinear contribution in the electric field \cite{bloembergen1982}. The relation between the electric polarization and the electric field strength is generally given as \cite{boyd2019}; 
\begin{equation}
\begin{aligned}
P &=\varepsilon_{0} \chi^{(1)} E+\varepsilon_{0} \chi^{(2)} E^{2}+\varepsilon_{0} \chi^{(3)} E^{3}+\cdots \\
&=P_{L}+P_{N L}.
\end{aligned}
\end{equation}
The first term, $\varepsilon_{0} \chi^{(1)} E$, is the linear polarization, and the higher-order terms generate nonlinear polarization. $\chi^{(n)}$ ($n>1$) are the $n$th-order nonlinear susceptibility tensors. 
For example, $\chi^{(2)}$ is a third rank tensor and relates the second-order nonlinear polarization to the electric field intensity which is proportional to the square of electric field strength. Second-order effects consist of the sum frequency generation (SFG, where the output frequency is the sum of the two input frequencies), second harmonic generation (SHG, where the output frequency is twice the input frequency), linear electro-optic effect, optical parametric amplification, etc. \cite{franken1961,baumgartner1979optical,kurtz1967physical}. Common third-order effects include third harmonic generation, four-wave mixing, optical Kerr effect, two-photon absorption, etc. \cite{maker1965study,yariv1977amplified,maker1964intensity,kaiser1961two}. The study of nonlinear optics began in the 60s with the first experimental demonstration of SHG in 1961 \cite{franken1961}. Later, Bloembergen, Boyd, Shen, and co-workers formulated the basic principles of the topic \cite{armstrong1962interactions,shen1984principles,boyd2019}. These nonlinear effects are typically weak, hence, practical nonlinear optics require bulk materials orders of magnitude larger than the optical path length for the effect to be significant. Recent advent of new nonlinear effects in two-dimensional (2D) atomically thin crystals has shown 2$\sim$3 orders of magnitude larger values of normalized $\chi^{(2)}$, hence drawing interest from the community \cite{autere2018nonlinear,guo20192d}.

In particular, phase mismatch is an important issue in bulk nonlinear optics, when the total phase of the output is not equal to that of its inputs. For SHG, the phase mismatch is described as $\Delta k=2k_1-k_2=2k(\omega)-k(2\omega)$. The nonlinear output intensity without phase matching (i.e $\Delta k \neq 0$) is expressed as \cite{boyd2019}
\begin{equation}
I_{2\omega}(L)=\frac{2 \left[\chi^{(2)}\right]^2\omega_{3}^{2} I_{1}^2(0)}{n_{\omega}^2 n_{2\omega} \epsilon_{0} c^{3}} L^{2} \operatorname{sinc}^{2}\left(\frac{\Delta k L}{2}\right).
\end{equation}
\begin{figure}[t]
	\includegraphics[width=\linewidth]{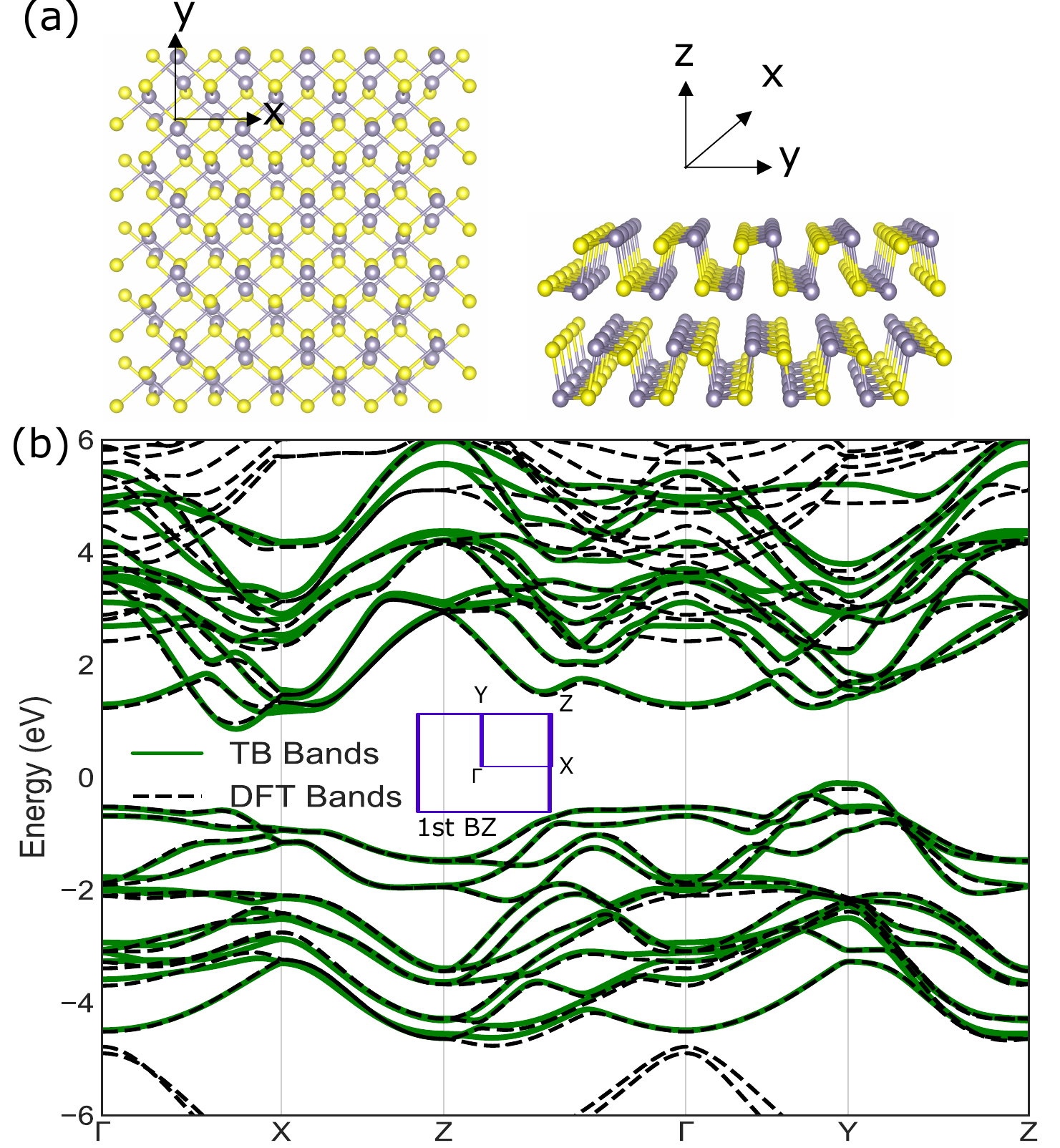}
	\caption{The top and side view of bilayer SnS atomic structure are shown in (a). Sn and S atoms are indicated by the grey and yellow balls, respectively. Panel (b) shows the band diagram of bilayer SnS, where solid (dashed) lines denote TB (DFT) bands. The rectangular first Brillouin zone is shown in the inset. There is a very good agreement between TB and DFT bands for the low-energy range of interest.}
	\label{fig1}
\end{figure}
where the intensity is proportional to the phase mismatch, $\Delta k$, input intensity $I_1$, length of the crystal $L$, and the SHG coefficient $\chi^{(2)}$. The phase mismatch factor is maximum when $\Delta k L=0$. For finite $\Delta k L$, the phase mismatch factor decreases exponentially and the output intensity is reduced drastically. 
To satisfy the phase matching condition ($\Delta k=0$), the refractive index of the material has to be equal at both the fundamental and second harmonic frequencies i.e $n(2\omega)=n(\omega)$, which is usually not achievable due to material dispersion. One way to solve this problem is to use birefringent crystals, where ordinary and extraordinary waves have different dispersion, and phase matching condition can be satisfied \cite{midwinter1965effects}. But in this case, the input and output are bound to be orthogonally polarized. Another technique for phase mismatch correction is called quasi-phase-matching (QPM) \cite{feng1980enhancement,yamada1993first}, which requires  complicated setup.

Hence, large $\chi^{(2)}$ and small $\Delta k$ are necessary for efficient nonlinear optics. In this regard, 2D materials can be advantageous over traditional bulk nonlinear materials. Typically reported $\chi^{(2)}$ in 2D materials such as MoS$_\text{2}$, WSe$_\text{2}$, h-BN, and GaSe \cite{wang2015nonlinear,autere2018nonlinear}, are 2$\sim$3 orders of magnitude larger than that in prototypical bulk materials (eg. LiNbO$_\text{3}$, Quartz, KDP, BBP, GaAs). For example, $\chi^{(2)}$ of MoS$_\text{2}$ was reported experimentally to be $5 \times 10^3 - 10^5$ pm/V in \cite{kumar2013second}, whereas $\chi^{(2)}$ of well-known nonlinear material LiNbO$_\text{3}$ is $\sim 60$ pm/V. The phase mismatch becomes negligible because of the atomic thickness of 2D layers  ($\sim$nm) compared to bulk materials ($\sim$mm) \cite{nikogosyan2006nonlinear,boyd2019}. Since nonlinear output intensity is also proportional to the device size ($L$), a further 2-3 orders of magnitude increment of $\chi^{(2)}$ in 2D materials is required to compensate for the reduction in device dimensions. In this work, we present an approach where such giant $\chi^{(2)}$ can be realized through double resonance effect. This is achieved by electrostatic tuning of the bandstructure, achieving triplet of nested bands which allows for resonance with both the $\omega$ and $2\omega$ transitions. 

In general, nested double resonant bands should be rare occurrence in pristine 2D materials. Hence, we narrowed our choices to 2D materials with quantum-well-like properties, whose electronic bandstructure are easily tunable with an out-of-plane electric field \cite{low2014tunable,low2014plasmons}, due to the strong interlayer coupling \cite{ozccelik2018tin}. We study SnS, a group IV transition metal monochalcogenide, which is isoelectronic with the  puckered honey-comb structure of black phosphorus (BP) \cite{gomes2015phosphorene} (see Figure \ref{fig1}). \textcolor{black}{Similar to BP, SnS has anisotropic optical properties (see linear optical properties of bilayer SnS in Supplementary Material \cite{supp}  Section A, also see the references therein \cite{low2014tunable,de2017multilayered,zhang2017infrared}). Unlike BP, which belongs to $\text{D}_{\text{2h}}$ point group, SnS has a reduced symmetry ($\text{C}_{\text{2v}}$ point group) and is non-centrosymmetric for odd-numbered layers  \cite{wang2017giant}. The even-numbered layers retain the inversion center, which can be broken by applying an out-of-plane electric field introduced by modifications of the onsite potentials of the constituent atoms. By tuning the applied perpendicular bias, the bandstructure of bilayer SnS, and therefore its SHG coefficients can be tuned.}

\section{Theory \& Approach}

In this work, we focus on the second-order nonlinearity, specifically the SHG. As the second order nonlinear susceptibility ($\chi^{(2)}(\omega_3;\omega_1,\omega_2)$) (where $\omega_1,\omega_2$ are input frequencies, and $\omega_3 = \omega_1 + \omega_2$ is the output frequency) is a third rank tensor, there are 27 different tensor elements for each combination of the input and output frequencies. The number of independent non-zero tensor elements is greatly reduced when symmetry operations are taken into account. Considering permutation symmetry, we can write
$\chi_{i j k}^{(2)}\left(2\omega ; \omega, \omega \right)=\chi_{i k j}^{(2)}\left(2\omega ; \omega, \omega\right)$,
which means that if one permutes the last two indices in $\chi_{i j k}^{(2)}$, the value will be the same, because the order of the input fields is not important. This reduces the maximum number of non-zero tensor elements from 27 to 18. Crystallographic symmetries can further reduce the number of non-zero tensor elements.
For example, there are only 7 non-zero independent tensor elements for point group $\text{C}_{\text{2v}}$ ($xzx, xxz, yyz, yzy, zxx, zyy, zzz$) \cite{boyd2019}.
According to Neumann's principle \cite{neumann1885vorlesungen}, any physical property of a crystal has to remain invariant after applying symmetry operations belonging to that crystal. However, in centrosymmetric systems, the existence of an inversion center requires polarization to be anti-symmetric with an external electric field. For second-order polarization, this is only satisfied when $\chi^{(2)}$ is zero. As a result, all the terms in $\chi^{(2)}$ vanish and these crystals cannot produce any bulk SHG response. It is possible to break the inversion  symmetry by applying external stimulus such as DC electric field \cite{seyler2015electrical,klein2017electric,shree2021interlayer}, DC current \cite{wu2012quantum}, strain \cite{rhim2015strain}, or structural engineering by patterning nanostructured arrays \cite{lan2016electrically,butet2015optical}. Thus, tunable non-zero SHG can in principle be achieved in centrosymmetric materials. 

\begin{figure}[htbp]
	\includegraphics[width=\linewidth]{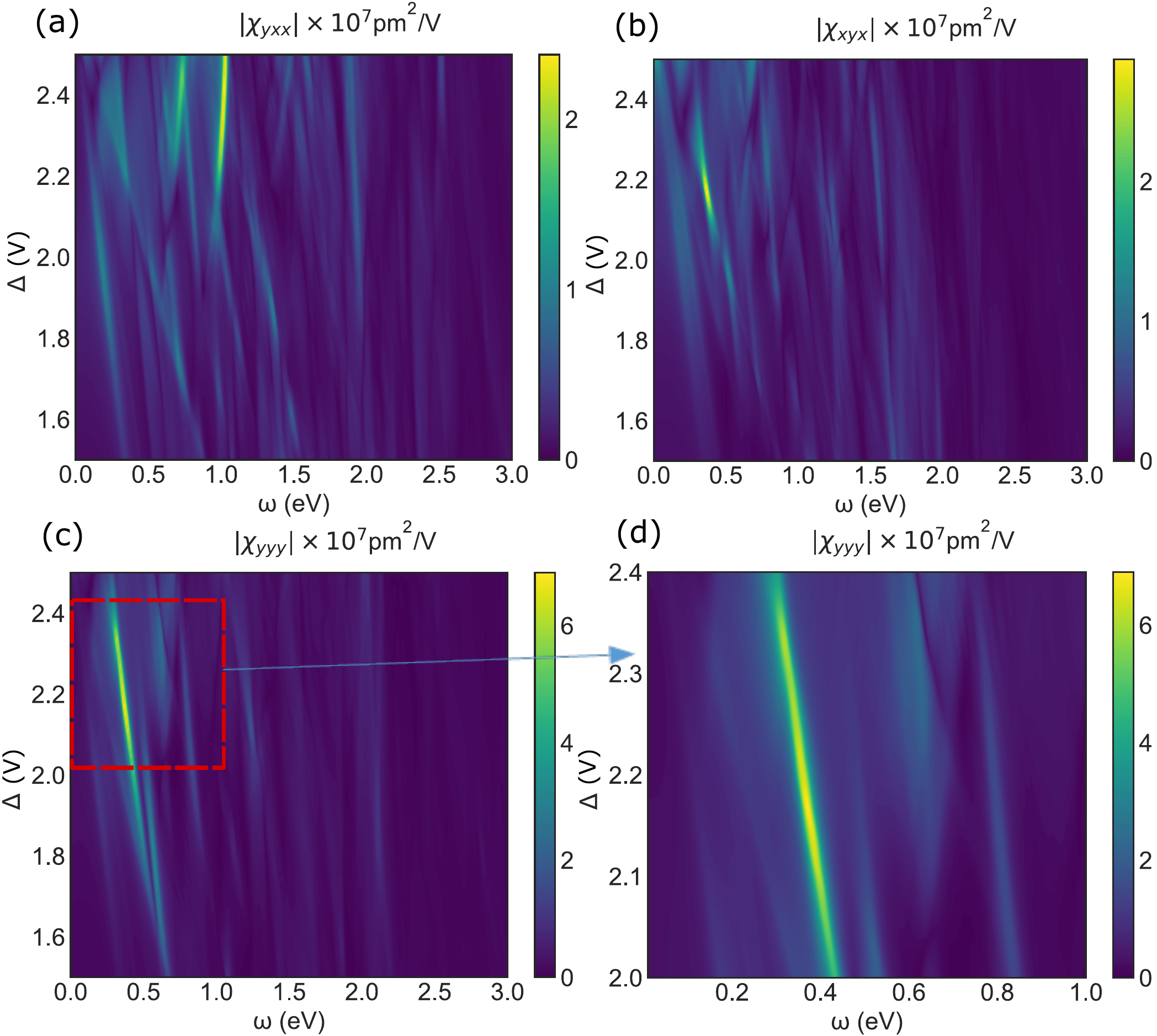}
	\caption{Three different components of sheet susceptibility $\chi^{(2)}$ are shown. Panels (a), (b), and (c) show $\chi_{yxx}$, $\chi_{xyx}$, and $\chi_{yyy}$ respectively for bilayer SnS as a function of applied bias $\Delta$ and frequency $\omega$. The maximum value of $\chi^{(2)}$ is achieved for $\chi_{yyy}$, $\sim 7 \times 10^7$ pm$^\text{2}$/V . The zoomed plot of $\chi_{yyy}$ is shown in (d) where a prominent peak occurs for $\Delta \sim 2.16$ V.  }
	\label{fig2}
\end{figure}

We use a tight-binding (TB) Hamiltonian generated by Wannier interpolation \cite{wang2006ab,yates2007spectral} from density functional theory (DFT) calculations. To construct a reliable TB model for multilayer SnS, we performed first-principles calculations to calibrate the effective Hamiltonian, using the Quantum Espresso package \cite{giannozzi2009quantum,giannozzi2017advanced}. Our parametrization procedure in this work is based on the formalism of Maximally Localized Wannier Functions (MLWFs) \cite{Marzari1997,Marzari2012} as implemented in the Wannier90 code \cite{Mostofi2008}. Ultrasoft Perdew-Burke-Ernzerhof (PBE) potential \cite{perdew1996generalized} was used to describe the exchange interactions with a kinetic energy cutoff of 60 Ry. A 12$\times$12$\times$1 Monkhorst- Pack \cite{monkhorst1976special} grid was used for the Brillouin Zone sampling for both the relaxation and static calculations. A vacuum thickness of 20 $\si{\angstrom}$ was introduced to avoid spurious interactions between adjacent images in the direction perpendicular to the 2D plane. 
Diagonalizing the TB Hamiltonian, we get the eigenenergies and eigenfunctions, which are used to calculate the SHG coefficient $\chi^{(2)}$ using the following equations. The total SHG coefficient consists of three terms \cite{hughes1996calculation}:  
\begin{widetext}
\begin{subequations}
	\begin{equation}
	\chi_{\mathrm{inter}}^{a b c}(2 \omega ; \omega, \omega)=\frac{e^{3}}{\hbar^{2}} \sum_{n m l} \int \frac{d \mathbf{k}}{4 \pi^{3}} \frac{r_{n m}^{a}\left\{r_{m l}^{b} r_{l n}^{c}\right\}}{\omega_{l n}-\omega_{m l}}\left\{\frac{2 f_{n m}}{\omega_{m n}-2 \omega}+\frac{f_{m l}}{\omega_{m l}-\omega}+\frac{f_{l n}}{\omega_{l n}-\omega}\right\},
	\label{eq:chi_inter}
	\end{equation}
	
	\begin{equation}\begin{aligned}
	\chi_{\mathrm {intra }}^{a b c}(2 \omega, \omega, \omega)=& \frac{e^{3}}{\hbar^{2}} \int \frac{d \mathbf{k}}{4 \pi^{3}}\left[\sum_{n m l} \omega_{m n} r_{n m}^{a}\left\{r_{m l}^{b} r_{l n}^{c}\right\}\left\{\frac{f_{n l}}{\omega_{l n}^{2}\left(\omega_{l n}-\omega\right)}-\frac{f_{l m}}{\omega_{m l}^{2}\left(\omega_{m l}-\omega\right)}\right\}\right.\\
	&\left.-8 i \sum_{n m} \frac{f_{n m} r_{n m}^{a}\left\{\Delta_{m n}^{b} r_{m n}^{c}\right\}}{\omega_{m n}^{2}\left(\omega_{m n}-2 \omega\right)}+2 \sum_{n m l} \frac{f_{n m} r_{n m}^{a}\left\{r_{m l}^{b} r_{l n}^{c}\right\}\left(\omega_{m l}-\omega_{l n}\right)}{\omega_{m n}^{2}\left(\omega_{m n}-2 \omega\right)}\right],
	\end{aligned}\end{equation}
	
	\begin{equation}
	\begin{aligned}
	\chi_{\mathrm{mod}}^{a b c}(2 \omega, \omega, \omega)&=\frac{e^{3}}{2 \hbar^{2}} \int \frac{d \mathbf{k}}{4 \pi^{3}}\left[\sum_{n m l} \frac{f_{n m}}{\omega_{m n}^{2}\left(\omega_{m n}-\omega\right)}\left\{\omega_{n l} r_{l m}^{a}\left\{r_{m n}^{b} r_{n l}^{c}\right\}-\omega_{l m} r_{n l}^{a}\left\{r_{l m}^{b} r_{m n}^{c}\right\}\right\}\right.\\
	&\left.-i \sum_{n m} \frac{f_{n m} r_{n m}^{a}\left\{r_{m n}^{b} \Delta_{m n}^{c}\right\}}{\omega_{m n}^{2}\left(\omega_{m n}-\omega\right)}\right].
	\end{aligned}
	\end{equation}
\end{subequations}
\end{widetext}
\begin{figure*}[t]
	\includegraphics[width=0.8\linewidth]{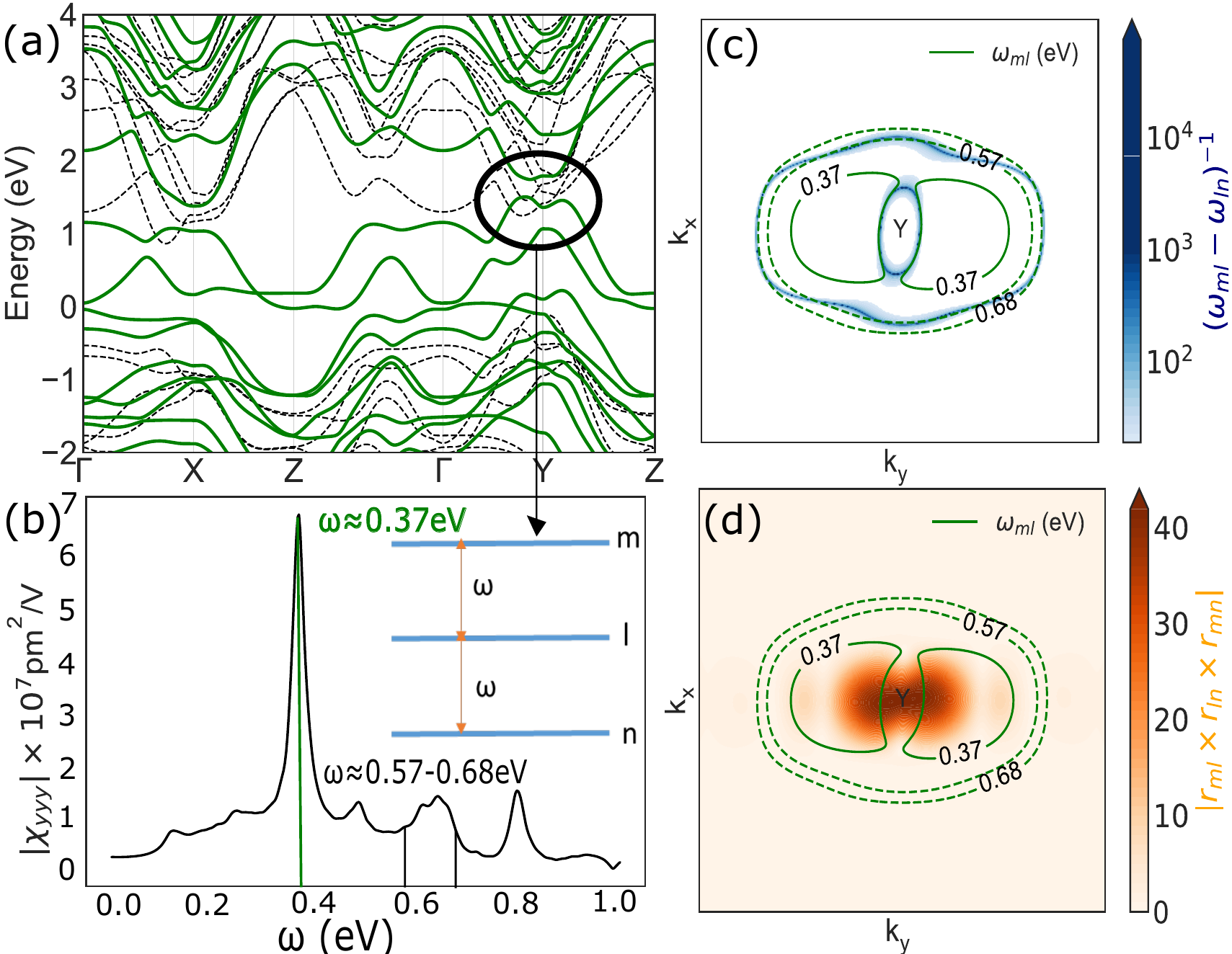}
	\caption{Schematic of double resonance in bilayer Sns. (a) Bandstructure of bilayer SnS with/without perpendicular bias of $\Delta=2.16$ V. The solid (dashed) lines show the biased (unbiased) bandstructure. Near point Y (circled), top two valence bands ($l$, $n$) and bottom conduction band ($m$) appear to be nested. $\chi_{yyy}$ for  $\Delta=2.16$ V is shown in (b). The prominent peak is located at $\omega \simeq 0.37$ eV. In panels (c,d), Brillouin zone contour plots are shown around point Y. The blue contours in (c) show the inverse of energy difference ($\omega_{ml}-\omega_{ln}$) on log scale, while the orange color in (d) show the magnitude of matrix element products between the bands. The regions where maximum of these two features overlap are where $\chi^{(2)}$ would be maximum, and the frequency contour that coincides with this region is $\omega_{ml}=0.37$ eV, which gives peak SHG frequency of 0.74 eV.}
	\label{fig3}
\end{figure*}
Here, $\chi_{\mathrm{inter}}^{a b c}$ is the interband contribution, $\chi_{\mathrm{intra}}^{a b c}$ is a modification due to intraband motion, and  $\chi_{\mathrm{mod}}^{a b c}$ is a modulation of intraband motion by interband polarization energy. $r_{n m}^{a}$ are the matrix elements of position operator between bands $n$, $m$ along the $a$ direction, $f_{nm}=f_n-f_m$ is the difference of Fermi-Dirac factors between bands $n$, $m$; $\omega_{n m}$ is the energy difference between bands $n$, $m$. The band indices $n$, $m$, $l$ runs over all bands. $\left\{r_{m l}^{b} r_{l n}^{c}\right\}=\frac{1}{2}\left(r_{m l}^{b} r_{l n}^{c}+r_{m l}^{c} r_{l n}^{b}\right)$ is used to ensure intrinsic permutation symmetry. $\Delta_{m n}^{b}=v_{mm}^b-v_{nn}^b$, where $v_{nm}^b$ is the velocity matrix element in the $b$ direction given by $v_{nm}^b=i\omega_{nm}r_{nm}^b$. 

From an inspection of the terms in the susceptibility equations, we can formulate a criterion for maximizing $\chi^{(2)}$. For example in Eqn. \eqref{eq:chi_inter}, there are $\omega$ and $2\omega$ terms in the denominator. When we have three bands $m$, $l$, $n$ satisfying the condition $\omega_m-\omega_l=\omega_l-\omega_n$, both the $\omega$ and $2\omega$ terms become resonant. The scenario where the bands are equidistant over some area in the 2D k-space (Brillouin zone) is called ``band nesting'', and the so-called ``double resonance" condition \cite{nevou2006intersubband} is satisfied, rendering significant enhancement of $\chi^{(2)}$. 

\section{Results \& Discussion}
\begin{figure*}[t]
	\begin{center}
		\includegraphics[width=0.8\linewidth]{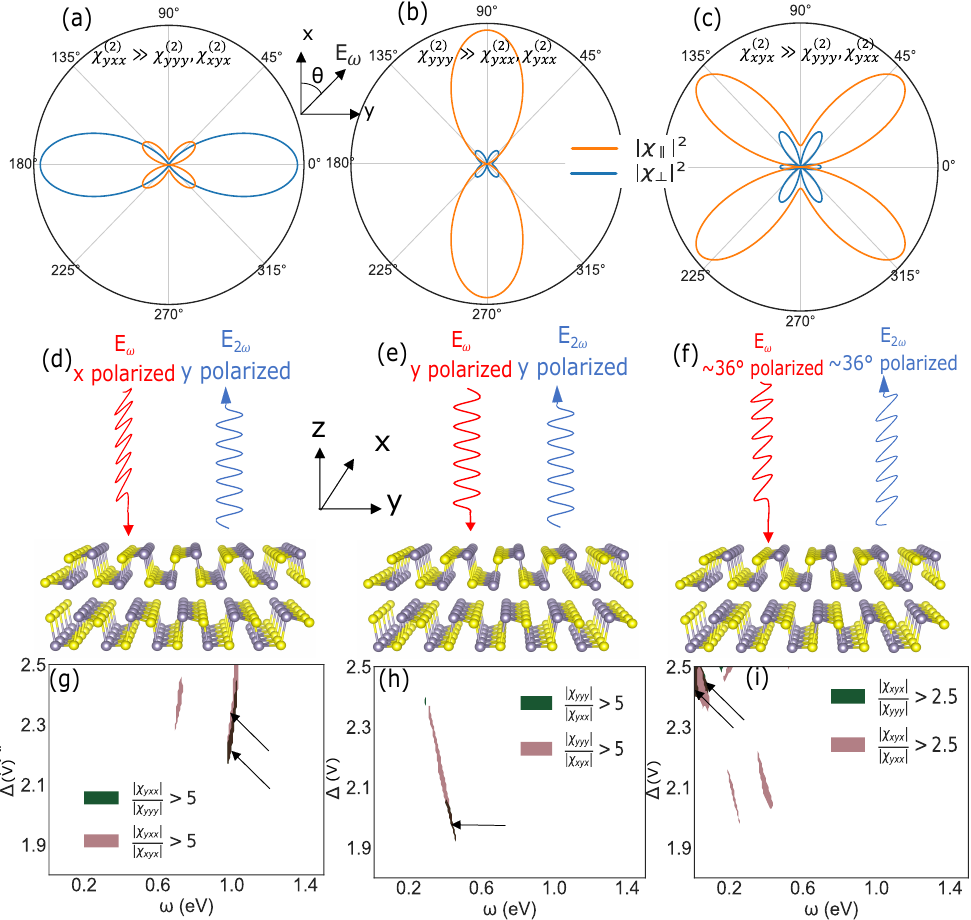}
		\caption{(a), (b), (c) show angle resolved polarization components of  $\chi^{(2)}$ $(|\chi_{\bot}|^2$, $|\chi_{\parallel}|^2)$ as a function of polarization angle $\theta$ with respect to $x$ axis, when one component is much larger than the other two. The effect of having one dominant $\chi^{(2)}$ component is shown in (d), (e), and (f) respectively. For example, in (a) where $\chi^{(2)}_{yxx}$ is dominant, if input is $x$ polarized ($\theta=0$) the perpendicular SHG component (i.e $y$ polarized) is maximum, so in (d) the SHG output will be $y$ polarized. Similarly when $\chi^{(2)}_{yyy}$ is dominant (b,e), the SHG output will be $y$ polarized with $y$ polarized input and when $\chi^{(2)}_{xyx}$ is dominant (c,f), the SHG output will be maximized with a polarization angle of $\sim 36^{o} $ and for both purely $x$ and $y$ polarized light ($\theta=0, 90^{o}$) the output is much smaller. Panels (g-i) show possible regions in the ($\omega , \Delta$) space where these conditions can be satisfied. The two different colors show where one $\chi^{(2)}$ component is significantly larger (5 times in (g), (h); 2.5 times in (i); $\chi^{(2)}>0.5 \chi^{(2)}_{max}$) than the other components. The overlap regions (marked by black arrows) are where the conditions in panels (a-c) are satisfied.}
		\label{fig4}
	\end{center}
\end{figure*}
Figure \ref{fig1} shows the atomic (panel a) and electronic (panel b) structure of bilayer SnS. The layers are AB stacked, and $x$ and $y$ denote zigzag and armchair directions, respectively. The unit cell consists of 4 pairs of Sn and S atoms stacked in AB configuration, which is more favorable in energy than other configurations. The bandstructure of bilayer SnS is plotted in Figure \ref{fig1}(b). Our first-principles calculations result shows that it is an indirect gap semiconductor with the valence band maximum located in the highly symmetric Y point and the conduction band minimum located between $\Gamma$ and $X$ points. To construct a reliable TB model, the $p_x$, $p_y$, and $p_z$ orbitals of Sn and S are chosen. The bandstructure derived from the 24-orbital TB Hamiltonian shows that there is a very good agreement between DFT and TB bands, especially in the low-energy region of interest.
Since the bandstructure of bilayer SnS is well reproduced from the TB model, we can further study the effect of external electric field on the susceptibility. The calculated $\chi^{(2)}$ parameters in 2D (sheet) units are shown in Figure \ref{fig2} as a function of applied perpendicular bias $\Delta$ and frequency $\omega$. The temperature $T$ is set to be 300K, and electron relaxation parameter which is used to ``broaden" the frequency ($\omega \equiv \omega+i\Gamma$) is set to be 10 meV. The out-of-plane electrical bias is introduced by adding a diagonal matrix to the TB Hamiltonian, where each diagonal element is the onsite potential of the corresponding atom due to the electric field.
There are three  independent $\chi^{(2)}$ components for $\text{C}_{\text{2v}}$ point group symmetry in 2D: $\chi_{yyy}$, $\chi_{yxx}$ and $\chi_{xyx}=\chi_{xxy}$ \cite{wang2017giant}. Due to anisotropy, each component has a different maximum amplitude for different bias and frequency. \textcolor{black}{ At higher biases the susceptibility becomes highly nonlinear as function of the bias.} The overall maximum amplitude of $\chi^{(2)}$, $\chi^{(2)}_{max}$, is about $\sim 7 \times 10^7$ pm$^\text{2}$/V for the $\chi_{yyy}$ component, which corresponds to $\sim8 \times 10^4$ pm/V in equivalent bulk units.  This value of $\chi^{(2)}$ is $\sim3$ orders of magnitude larger than typically reported theoretical values of nonlinear 2D materials. The experimental values of $\chi^{(2)}$ in 2D materials vary a lot across experiments. For example, the value for $\text{MoS}_2$ ranges from 1.2 to 10$^\text{5}$ pm/V, which is a large variation. Factors like substrate and sample preparation process, defects, and excitonic effects can influence the measurement \cite{kumar2013second,malard2013observation}. The theoretical values calculated within DFT lie in the middle of this range, \cite{wang2015nonlinear} ($\sim10^3$ pm/V). Thus, it is more meaningful to compare our calculations with the theoretical values. \textcolor{black}{To draw further comparison, this vale of $\sim8 \times 10^4$ pm/V is higher than reported values in 2D monochalcogenides (GeS, GeSe, SnS, SnSe) monolayers which goes up to $10^4$ pm/V \cite{zhang2017anisotropic}.} The required perpendicular bias for $\chi^{(2)}_{max}$ is $\Delta \sim 2.16$ V, which  corresponds to an electric field of $2.5$ V/nm, which is smaller than the breakdown electric field of Si$\text{O}_{\text{2}}$ \textcolor{black}{thin films (3-4 V/nm)} \cite{sire2007statistics}. Accordingly, the corresponding SHG frequency is $0.74$ eV.

The SHG enhancement by double resonance is explained in Figure \ref{fig3}. Fig. \ref{fig3}(a) shows the modified bandstructure of bilayer SnS at bias that yields $\chi^{(2)}_{max}$, in conjunction with the unbiased bandstructure.  A cutline of $\chi_{yyy}$ along $\Delta=2.16$ V shows the prominent peak at $\omega \simeq 0.37$ eV (Fig. \ref{fig3}(b)). To explain the double resonance criterion, we identify a set of three bands in the vicinity of $Y$ point in the Brillouin zone, which is highlighted by the black circle. The relevant band parameters around $Y$ point are plotted in Figures \ref{fig3}(c) and \ref{fig3}(d). The quantity $(\omega_{ml}-\omega_{ln})^{-1}$, which quantifies the double resonance condition, is showed in log scale by the blue contours in Fig. \ref{fig3}(c). The orange color in Fig. \ref{fig3}(d) shows the magnitude of the matrix element product between bands $m$, $l$, and $n$ in log scale. In most cases, $\chi^{(2)}$ would be optimal when both these quantities are maximized. To determine the frequency ($\omega_{ml}=\omega$) of this $\chi^{(2)}_{max}$, we plot contour lines of $\omega_{ml}$ in the region of interest. Our result shows that $\chi^{(2)}$ is maximum at $\omega=0.37$ eV, where the  solid contour line coincides with the hot spots of these quantities in the momentum space, hence $\chi^{(2)}$ is maximum at this frequency. Although some of the dashed contours at $\omega \simeq 0.57-0.68$ eV coincide with the double resonance regions, the matrix elements product along those contours is at least 2 orders of magnitude smaller, which negates the enhancement gained from double resonance. Indeed, we can observe small features in $\chi^{(2)}$ spectra around these frequencies (Fig. \ref{fig3}(b)). The rest of the smaller features seen in $\chi^{(2)}$ are most likely originating from other possible nested trios of bands, albeit with a lesser degree of nesting or smaller matrix element product.  
\textcolor{black}{SHG response depends on $\Delta$ in a highly nonlinear fashion. However, for $\Delta \simeq 0$, SHG scales linearly \cite{shree2021interlayer}, except near resonant peaks (see Supplementary Material \cite{supp} section B).}

Next we turn to the tunability of polarization anisotropy of bilayer SnS. For each crystallographic point group, the effective $\chi^{(2)}$ at an angle of $\phi$ can be derived from the input field polarization $\theta$ and the susceptibility tensor \cite{shen1984principles}:
\[\chi_{\phi}^{(2)}=\mathbf{e_{\phi}} \cdot \mathbf{\chi^{(2)}} : \mathbf{e_{\theta}} \otimes \mathbf{e_{\theta}}.  \]
Here $\chi_{\phi}^{(2)}$ is the effective SHG coefficient along the $\phi$ direction, $\theta$ is the input field polarization angle and $\mathbf{e_{\phi}}$, $\mathbf{e_{\theta}}$ are the corresponding unit vectors. The SHG output intensity, $I^{2\omega}_{\phi} \propto \left|\chi_{\phi}^{(2)}\right|^2$.  For the $\text{C}_{\text{2v}}$ point group, the angle-resolved SHG susceptibilities are given by \cite{wang2017giant}:
\begin{equation}
\begin{array}{l}
\chi_{\|}^{(2)}\left(\theta\right)=\left(\chi_{x y x}^{(2)}+\chi_{y x x}^{(2)}\right) \sin \theta \cos ^{2} \theta+\chi_{y y y}^{(2)} \sin ^{3} \theta, \\
\chi_{\perp}^{(2)}\left(\theta\right)=\chi_{y x x}^{(2)} \cos ^{3} \theta+\left(\chi_{y y y}^{(2)}-\chi_{x y x}^{(2)}\right) \cos \theta \sin ^{2} \theta.
\end{array}
\end{equation}
Here $\chi_{\|(\perp)}^{(2)}$ are the parallel (perpendicular) polarization component of SHG coefficient with respect to the input electric field ($E_{\omega}$).

The tunability of SHG components suggests the interesting scenarios where one of the components is dominant over the other two. The angle-resolved polarization components of $\chi^{(2)}$ for these three cases are shown 
in Figure \ref{fig4}(a-c), and the corresponding input vs output polarization are illustrated in panel (d-f). For example, when  $\chi_{yxx}^{(2)}$ is dominant, the perpendicular SHG component, hence the intensity of SHG output, is maximum when $\theta=0$. This means an x-polarized input will result in a y-polarized output, while the output intensity for the other polarization directions will be significantly weaker. Similarly, for dominant $\chi_{yyy}^{(2)}$, when the input is  y-polarized, the output will be maximized and y-polarized. For dominant $\chi_{xyx}^{(2)}$, neither of the x and y-polarized inputs are converted to strong SHG output. However, the output is maximized with a polarization angle of $\sim 36^{\circ}$. In Fig. \ref{fig4}(g-i), we analyze the tuning parameter space ($\omega$, $\Delta$) for each of the cases (a-c). The two colors show regions where the SHG component is larger by a factor of $K$ ($K=5$ in panels 4(g) and 4(h); 2.5 in panel 4(i)). Herein, only the regions where $\chi^{(2)}>0.5 \chi^{(2)}_{max}$ are considered as we are interested in the strong SHG response. The overlap of the two regions indicated by black arrows shows where the component is dominant, and the input-output polarization selectivity is effective. Thus, the tunability of individual SHG parameters allow enhancement of the polarization anisotropy. 

\textcolor{black}{In this work, the independent particle approximation (IPA) has been used, and many body effects such as quasi-particle self-energy corrections and excitonic effects are not considered. However, in 2D systems, the effects of many-body interactions are expected to be pronounced due to reduced screening effect and quantum confinement \cite{wang2015nonlinear}. Simple DFT calculations with IPA typically underestimate the bandgap measured in optical experiments \cite{huser2013quasiparticle,shishkin2007self}. Quasi-particle bandstructure using $\text{GW}_{\text{0}}$ approximation can be calculated by many-body perturbation theory to determine the GW bandgap correction \cite{shishkin2006implementation,wang2017two}. We initially constructed a bilayer-SnS TB model via $\text{GW}_{\text{0}}$ calculations (see Supplementary \cite{supp} section C, see also the references therein \cite{Perdew1997,Marzari2012,Kresse1999}), to get a more accurate description of quasi-particle energy.  But the agreement between TB bands and $\text{GW}_{\text{0}}$ bands was poor with this model. Thus, we decided to use a more complex 24-band TB model that can fit the DFT bands more accurately (see Fig. \ref{fig1}(b)).}
	
\textcolor{black}{For this class of 2D materials, the DFT and GW bandstructures are similar \cite{wang2017giant}. To include excitonic effects, first principles calculations in the Bethe-Salpeter exciton approach \cite{salpeter1951relativistic,onida2002electronic} is extremely computationally challenging, as it requires dense k-point grid with thousands of points, and a high number of bands, to achieve a reliable agreement with experiment \cite{qiu2013optical,wang2015nonlinear}. For nonlinear susceptibility calculations, the complex expressions involving dipole matrix elements of position operators that require non-local and frequency dependent Hamiltonians, make the calculations even more difficult \cite{wang2015nonlinear,wang2017giant}.}

\textcolor{black}{Furthermore, we investigate the effect of perpendicular bias on the bandstructure, which requires repeating each of these computation-intensive first principles calculations hundreds of times. The double resonance effect we discussed here requires very fine tuning of the applied bias, rendering the calculations extremely prohibitive. Also, SHG response typically increases in the vicinity of exciton peaks when exciton effects are included, and away from the exciton peaks the effects are negligible \cite{trolle2014theory}. Thus, including the excitonic effects would not alter the main conclusion of our work, which is giant enhancement of SHG due to double-resonance achieved via band-tuning. In that regard, our DFT derived TB approach  provides a computationally feasible way to study the proposed effects, which captures the essential physics.}

\section{Conclusion}
The weak nonlinearity in bulk materials is a bottleneck towards realization of efficient nonlinear metasurfaces in low-dimensional systems, as the required intensity of input light is too high. There is also the problem of phase mismatch as discussed earlier. Achieving giant nonlinear response in 2D materials can solve both of these problems and facilitate the design of novel nonlinear devices such as frequency converters, optical modulators, sensors etc. in small volumes with high conversion efficiency.  
We proposed a fundamentally new approach to generating giant SHG responses in 2D materials. A tight-binding model for bilayer SnS was formulated to calculate SHG coefficients with a perpendicularly applied electric field. The results demonstrate a giant SHG susceptibility of bilayer SnS up to $\sim8 \times 10^4$ pm/V which is $\sim 3$ orders of magnitude larger than typically reported values. We attribute the large SHG enhancement to double resonance, achieved by band nesting. Moreover, the band tunability also allows modification of the existing polarization anisotropy of SHG components. \textcolor{black}{Recent studies \cite{zhu2021efficient,moqbel2022wavelength} have reported synthesis and measurement of strong anisotropic and polarization dependent SHG in few layer SnS films.} By patterning 2D layers to create plasmonic structures for strong light-matter interaction, further enhancement of nonlinear intensity can be achieved. Furthermore, the polarization anisotropy of SHG can be used to design electrically tunable novel nonlinear polarization-sensitive devices such as photodetectors, synaptic devices, polarization converters and switches \cite{wang2019recent,tian2016anisotropic,zhang2017anisotropic}. 

\section*{Author Contributions}
S.R.B. and T.L. contributed to the conceptualization of the key ideas, with D.R.C.'s participation. S.R.B. prepared the computational package for calculation of SHG coefficients, performed the SHG coefficient calculations, and wrote the original draft of the manuscript with contributions from J.Y. and S.Y. J.Y. performed the first-principles and Wannierization calculations with supervision from S.Y. S.R.B., J.Y., Z.W., D.R.C., C.Z., S.Y., and T.L. contributed to the reviewing and editing of the original draft, and participated in discussions of key results. T.L. supervised the project.

\noindent $^\dagger$S.R.B. and J.Y. contributed equally to this work.

\noindent Notes: The authors declare no competing financial interest.

\begin{acknowledgments}
S.R.B. and T.L. acknowledge partial funding support from NSF DMREF-1921629. J.Y., Z.W., and S.Y. acknowledge support from the National Key R\&D Program of China (Grant No. 2018FYA0305800), the program for professor of special appointment (Eastern Scholar) at Shanghai Institutions of higher learning and by Dutch Science Foundation NWO/FOM under Grant No. 16PR1024. J.Y., Z.W., and S.Y. also gratefully acknowledge support by the Netherlands National Computing Facilities Foundation (NCF) with funding from the Netherlands Organization for Scientific Research (NWO). S.R.B. and T.L also thank Andrei Nemilentsau for useful discussions.
\end{acknowledgments}



\bibliography{nonlinear_manuscript}

\end{document}